\documentclass[prl,showpacs,twocolumn,floats]{revtex4}

\usepackage{latexsym}
\usepackage[dvips]{graphics}

\begin{document}

\title{Universal measurement apparatus controlled by quantum software}

\author{Jarom\'{\i}r Fiur\'{a}\v{s}ek}
\author{Miloslav Du\v{s}ek}
\author{Radim Filip}

\affiliation{Department of Optics, Palack\'y University,
     17.~listopadu 50, 772\,00 Olomouc, Czech~Republic}

\begin{abstract}
We propose a quantum device that can approximate any projective
measurement on a qubit. The desired measurement basis is selected by the
quantum state of a ``program register''. The device is optimized with
respect to maximal average fidelity (assuming uniform distribution of
measurement bases). An interesting result is that if one uses two qubits in
the same state as a program the average fidelity is higher than if
he/she takes the second program qubit in the orthogonal state (with
respect to the first one). The average information obtainable by the
proposed measurements is also calculated and it is shown that it can get
different values even if the average fidelity stays constant. Possible
experimental realization of the simplest proposed device is presented.
\end{abstract}


\maketitle



Programmable quantum ``multimeters'' are devices that can realize any
desired generalized quantum measurement from a chosen set (either
exactly or approximately) \cite{DuBu}. Their main feature is that the
particular positive operator valued measure (POVM) is selected by the
quantum state of a ``program register'' (quantum software). In this
sense they are analogous to universal quantum processors
\cite{Nielsen97,Vidal00,Hillery02}. Quantum multimeters could play an
important role in quantum state estimation and quantum information
processing.

In this paper, we will describe a programable quantum device that can
approximately accomplish {\em any} projective von Neumann measurement on
a single qubit. Since it is impossible to encode an arbitrary unitary
operation (acting on a finite-dimensional Hilbert space) into a state
of a finite-dimensional quantum system \cite{Nielsen97} it is also
impossible to encode arbitrary projective measurement on a qubit into
such a state \cite{DuBu}. However, it is still possible to encode POVM's
that represent, in a certain sense, the best approximation of the
required projective measurements.

Suppose we would like to measure a qubit in the basis represented by two
orthogonal vectors $|\psi\rangle$ and $|\psi_\perp\rangle$. We want this
measurement basis be controlled by the quantum state of a program
register, $|\phi_p(\psi)\rangle$. An \emph{ideal} multimeter would map
the composite state of the measured system and the program register to
two fixed orthogonal pure states $| 0 \rangle$ and $| 1 \rangle$
according to
\begin{equation}
|\psi\rangle \otimes |\phi_{\rm p}(\psi)\rangle \rightarrow |0\rangle,
\qquad
|\psi_\perp\rangle \otimes |\phi_{\rm p}(\psi)\rangle \rightarrow |1\rangle.
\label{IDEAL}
\end{equation}
As mentioned above, such transformation cannot be implemented exactly.
Thus, our task is to find a realistic linear trace-preserving completely
positive (CP) map that represents the closest approximation to this
non-realistic map. We focus on the scenario when we always obtain one of
the two measurement results $|0\rangle$ or $|1\rangle$ but errors, i.e.
deviations from the ideal map (\ref{IDEAL}), may appear. Our aim is to
minimize the probability of error, i.e., we will maximize the
probability of the correct discrimination between states $|\psi\rangle$
and $|\psi_\perp\rangle$. In general we could optimize both the program
and the fixed transformation so as to optimally approximate map
(\ref{IDEAL}) for a given dimension of the program register. However,
this is an extremely hard problem that we will not attempt to solve in
its generality. Instead, we shall optimize the fixed transformation for
two natural choices of the program.

First we shall assume that the program register contains $N$ copies of
the state $|\psi\rangle$, $|\phi_{\rm p}(\psi) \rangle =
|\psi\rangle^{\otimes N}$. Our second choice of the program -- the
two-qubit state $|\psi\rangle|\psi_\perp\rangle$ -- is motivated by
recent results on optimum quantum state estimation. Gisin and Popescu
showed that the state of two orthogonal qubits
$|\psi\rangle|\psi_\perp\rangle$ encodes the information on the state
$|\psi\rangle$ better than state of two identical qubits
$|\psi\rangle|\psi\rangle$ \cite{Gisin99}. If we possess one copy of the
state $|\psi\rangle|\psi_\perp\rangle$ than we can estimate
$|\psi\rangle$ with fidelity ${\cal{F}}_\perp = (1+1/\sqrt{3})/2 \approx
0.7886$ which is slightly higher than the fidelity of optimum estimation
on one copy of two identical qubits, ${\cal{F}}_{||}=3/4$
\cite{Gisin99,Massar00}. One would thus expect that the state
$|\psi\rangle|\psi_\perp\rangle$ should also
give an advantage when used as a program of the multimeter. Rather
surprisingly, this is not the case and we shall see that it is better to
use two identical qubits $|\psi\rangle|\psi\rangle$.

In what follows we shall benefit from the isomorphism between CP maps
and bipartite positive semidefinite operators
\cite{Jamiolkowski72,Fiurasek01}. Let
$\cal{H}$ and $\cal{K}$ denote the Hilbert spaces of input and output
states, respectively. Choose basis $|i\rangle$ in  $\cal{H}$, define a
maximally entangled state $\sum_i |i\rangle |i\rangle$ on
${\cal{H}}^{\otimes 2}$ and apply the CP map to one part of this state.
The density matrix $\chi$ of the resulting state on Hilbert space
${\cal{H}}\otimes \cal{K}$ represents the CP map and the  relation
between input and output density matrices reads \cite{Fiurasek01}
\begin{equation}
\rho_{\rm out}= {\rm Tr}_{\rm in}[\chi \rho_{\rm in}^T
\otimes \openone_{\rm out}]
\label{INOUT}
\end{equation}
where $T$ stands for the transposition in the basis $|i\rangle$ and $\openone$
denotes an identity operator. The CP
map is trace preserving if the positive semidefinite operator $\chi$
satisfies the condition ${\rm Tr}_{\rm out}[\chi]=\openone_{\rm in}$.

Let us define the fidelity $F(\psi)$ of our multimeter projecting onto
states $|\psi\rangle$ and $|\psi_\perp\rangle$ as the probability that a
correct measurement result will be obtained when we send the states
$|\psi\rangle$ or $|\psi_\perp\rangle$ to the input randomly each with
probability one half. This probability can be interpreted as a success
rate of the discrimination between two orthogonal states $|\psi\rangle$
and $|\psi_\perp\rangle$. Assuming the program state to be
$|\psi\rangle^{\otimes N}$, the two relevant input states of the
multimeter read
\begin{equation}
|\Psi\rangle=|\psi\rangle\otimes |\psi\rangle^{\otimes N},
\qquad
|\Psi_{\perp}\rangle =
|\psi_\perp\rangle \otimes|\psi\rangle^{\otimes N}.
\label{PSI}
\end{equation}
The input Hilbert space of the multimeter is a tensor product of the
Hilbert space of signal qubit ${\cal{H}}_{\rm s}$ and symmetric
(bosonic) subspace ${\cal{H}}_+^N$ of the Hilbert space of $N$ qubits,
${\cal{H}}={\cal{H}}_{\rm s} \otimes{\cal{H}}_+^N$ and ${\rm
dim}{\cal{H}}=2(N+1)$. A trace preserving completely positive map $\chi$
transforms the input states onto the states of a single output qubit
that is subsequently measured in the computational basis. The outcome
$|0\rangle$ corresponds to the projection onto $|\psi\rangle$ while
$|1\rangle$ is associated with the projection onto $|\psi_\perp\rangle$.
Making use of the input-output relation (\ref{INOUT}) we have,
\begin{eqnarray}
F(\psi)&=&\frac{1}{2}{\rm Tr} \Bigl[ \chi
(|\Psi\rangle\langle \Psi|)^T \otimes |0\rangle \langle 0| \Bigr]
\nonumber \\
&+&\frac{1}{2}{\rm Tr} \Bigl[ \chi \,
(|\Psi_\perp\rangle\langle \Psi_\perp|)^T \otimes |1\rangle \langle 1| \Bigr].
\label{FPSI}
\end{eqnarray}
The figure of merit that we would like to maximize is the mean fidelity
obtained on averaging  $F(\psi)$ over all pure qubit states
$|\psi\rangle$, i.e., over the surface of the Bloch sphere,
\begin{equation}
F=\int_\psi d\psi \, F(\psi)= \rm Tr[R \chi].
\label{F}
\end{equation}
The positive semidefinite operator $R$ reads
\begin{equation}
R=R_+ \otimes |0\rangle \langle 0|+ R_{-} \otimes |1 \rangle \langle 1|,
\label{R}
\end{equation}
where the operators $R_+$ and $R_-$ acting on the input Hilbert space
$\cal{H}$ are given by integrals
\begin{equation}
R_{+}^T=\frac{1}{2}\int_\psi d\psi \, |\Psi\rangle\langle\Psi|,
\qquad
R_{-}^T=\frac{1}{2}\int_\psi d\psi  \, |\Psi_\perp\rangle\langle\Psi_\perp|.
\label{RPM}
\end{equation}
A straightforward calculation reveals that $R_+$ is proportional to the
projector onto symmetric subspace ${\cal{H}}_+^{(N+1)}$ of the Hilbert
space of $N+1$ qubits,
\begin{equation}
R_{+}= \frac{1}{2(N+2)} \Pi_{+}^{(N+1)}.
\label{RP}
\end{equation}
Furthermore, the sum of operators $R_+$ and $R_-$ is proportional to the
identity operator on the input Hilbert space,
$R_{+}+R_{-}= \openone_{\rm in}/[2(N+1)] $.
Thus we immediately have
\begin{equation}
R_{-} = \frac{1}{2(N+1)}\openone_{\rm in} - \frac{1}{2(N+2)} \Pi_{+}^{(N+1)} .
\label{RM}
\end{equation}

The determination of the optimum CP map amounts to maximization of the
linear function (\ref{F}) under the constraints $\chi\geq 0$ and ${\rm
Tr}_{\rm out}[\chi]=\openone_{\rm in}$. The optimum CP map that maximizes the
mean fidelity (\ref{F}) must satisfy the extremal equations
\cite{Fiurasek01,Audenaert01}
\begin{eqnarray}
(\lambda\otimes \openone_{\rm out} -R) \chi = 0,
\label{EXTREMALA}
 \\
\lambda \otimes \openone_{\rm out} -R \geq 0, \quad
\label{EXTREMALB}
\end{eqnarray}
where $\lambda$ is positive definite operator on the input Hilbert
space. Notice that the extremal equations (\ref{EXTREMALA}) and
(\ref{EXTREMALB}) resemble the Helstrom equations for optimum POVM that
maximizes the success rate in ambiguous quantum state discrimination
\cite{helst}. One can prove that if both Eqs.~(\ref{EXTREMALA}) and
(\ref{EXTREMALB}) are satisfied, then $\chi$ is indeed optimum CP map
and $F$ attains its global maximum on the convex set of trace-preserving
CP maps \cite{Audenaert01,Fiurasek01b}.

The extremal equations can be very efficiently solved numerically by
means of repeated iterations \cite{Fiurasek01}.
From the numerical results, we were able to conjecture the optimum CP map,
\begin{equation}
\chi = \Pi_+^{(N+1)} \otimes |0\rangle\langle 0| +
\Pi_{-}^{(N+1)} \otimes |1\rangle \langle 1|.
\label{CHIOPT}
\end{equation}
where $\Pi_-^{(N+1)}=\openone_{\rm in}-\Pi_{+}^{(N+1)}$ is a projector
onto subspace orthogonal to the symmetric subspace of $N+1$ qubits.
On inserting the
expressions (\ref{R}) and (\ref{CHIOPT}) into Eq.~(\ref{F}), we obtain
the mean fidelity
\begin{equation}
F= \frac{2N+1}{2N+2}.
\label{FMAX}
\end{equation}

Prior to analyzing the properties of the CP map (\ref{CHIOPT}) in
detail, let us prove its optimality. Taking into account the
trace-preservation condition ${\rm Tr}_{\rm out} [\chi] =\openone_{\rm in}$, we
find from Eq.~(\ref{EXTREMALA}) that $\lambda={\rm Tr}_{\rm out}[R
\chi]$. After some algebra we arrive at
\begin{equation}
\lambda= \frac{1}{2(N+1)}\openone_{\rm in}
         - \frac{1}{2(N+1)(N+2)} \Pi_+^{(N+1)}.
\label{LAMBDA}
\end{equation}
It is easy to check that the first extremal equation (\ref{EXTREMALA})
is satisfied. The inequality (\ref{EXTREMALB}) splits into two
independent inequalities for operators acting on input Hilbert space,
\[
A_1 = \lambda - R_{-} \geq 0, \qquad
A_2 = \lambda - R_{+} \geq 0.
\]
A straightforward calculation yields explicit formulas for $A_1$ and
$A_2$,
\begin{eqnarray}
A_1 &=&\frac{N}{2(N+1)(N+2)}\, \Pi_+^{(N+1)}, \nonumber \\
A_2 &=& \frac{1}{2(N+1)} \, \Pi_{-}^{(N+1)}.
\label{AEXP}
\end{eqnarray}
Since both these operators are positive semidefinite, the condition
(\ref{EXTREMALB}) is satisfied. This concludes our proof of the
optimality of the CP map (\ref{CHIOPT}).

The structure of the optimum CP map (\ref{CHIOPT}) indicates that this
map is a joint generalized quantum measurement on the signal qubit and
$N$ program qubits and the corresponding POVM has two elements
$\Pi_{+}^{(N+1)}$ and $\Pi_{-}^{(N+1)}$. We can now determine the effective
POVM carried out on the signal qubit,
\begin{eqnarray}
\Pi_{||} &=& {\rm Tr}_{\rm p} \left[ \openone_{\rm s} \otimes
(|\psi\rangle \langle\psi|)^{\otimes N}
\, \Pi_{+}^{(N+1)} \right],
\nonumber \\
\Pi_{\perp}&=& {\rm Tr}_{\rm p} \left[ \openone_{\rm s} \otimes
(|\psi\rangle \langle \psi|)^{\otimes N} \, \Pi_{-}^{(N+1)} \right],
\label{PROJ}
\end{eqnarray}
where  ${\rm Tr}_{\rm p}$ denotes trace over the program qubits. The outcome
$\Pi_{\perp}$ cannot occur if the input state is $|\psi\rangle$ because
the input state $|\Psi\rangle$ belongs to the
symmetric subspace of $N+1$ qubits and $\Pi_{||}$ clicks with certainty.
Hence the POVM element $\Pi_{\perp}$ must be proportional to the
projector $|\psi_\perp\rangle\langle \psi_{\perp}|$. Since the sum of
POVM elements  (\ref{PROJ}) is an identity operator, we have the
following ansatz,
\begin{eqnarray}
\Pi_{||} &=& |\psi\rangle\langle \psi| + (1-p)|\psi_\perp\rangle\langle
\psi_\perp|,
\nonumber \\
\Pi_{\perp} &=& p \, |\psi_\perp\rangle \langle \psi_\perp |.
\label{PROJGUESS}
\end{eqnarray}
The probability $p$ that $\Pi_\perp$ clicks when the input state is
$|\psi_\perp\rangle$ is given by $p=\langle \Psi_\perp|
\Pi_-^{(N+1)}|\Psi_\perp\rangle$. After some algebra we get $p=N/(N+1)$
and the effective POVM representing our universal multimeter reads
\begin{eqnarray}
\Pi_{||}&=& \frac{1}{N+1} \openone + \frac{N}{N+1}|\psi\rangle\langle \psi|,
\nonumber \\
\Pi_{\perp}&=& \frac{N}{N+1} |\psi_\perp\rangle \langle \psi_\perp |.
\label{PROJFINAL}
\end{eqnarray}
Notice that the POVM (\ref{PROJFINAL}) is asymmetric, which reflects the
asymmetry of the program register. Furthermore, the fidelity $F(\psi)$
is independent of $\psi$ and equal to the mean fidelity (\ref{FMAX}). In
the limit of infinitely large program register $(N\rightarrow \infty)$,
the POVM (\ref{PROJFINAL}) approaches the ideal projective measurement.

We now turn our attention to the program $|\psi\rangle
|\psi_\perp\rangle$. The optimum CP map for this program can be found
following the same procedure as described above for the program
$|\psi\rangle^{\otimes N}$. Briefly, one has to calculate the operator
$R$ and solve extremal equations (\ref{EXTREMALA}) and
(\ref{EXTREMALB}). We will not give the details of calculations here and
only present the results.  Similarly as before, the optimum CP map is a
generalized measurement on the signal qubit and two program qubits. The
two elements of this three-qubit POVM read,
\begin{eqnarray}
\Pi_+ &=& \frac{1}{2} \Pi_{+}^{(3)}+|\phi_1\rangle\langle \phi_1|
+|\phi_2\rangle\langle \phi_2|,
\nonumber \\
\Pi_{-} &=& \openone_{3}- \Pi_{+},
\label{PPM}
\end{eqnarray}
where $\openone_3$ is an identity operator on Hilbert space
of three qubits and
\begin{eqnarray}
|\phi_1\rangle&=&\frac{1}{2\sqrt{3}}[
(\sqrt{3}+1)|0\rangle_{\rm s} |01\rangle_{\rm p}
-(\sqrt{3}-1)|0\rangle_{\rm s} |10\rangle_{\rm p}
\nonumber \\
&&-2\,|1\rangle_{\rm s} |00\rangle_{\rm p}],
\nonumber \\
|\phi_{2}\rangle&=&\frac{1}{2\sqrt{3}}[
(\sqrt{3}+1)|1\rangle_{\rm s} |10\rangle_{\rm p}
- (\sqrt{3}-1)|1\rangle_{\rm s} |01\rangle_{\rm p}
\nonumber \\
&&-2\,|0\rangle_{\rm s} |11\rangle_{\rm p}].
\label{FI}
\end{eqnarray}
Here the subscripts s and p label the states of signal and program qubits,
respectively. After some algebra, we find the effective POVM carried out
on the signal qubit,
\begin{eqnarray}
\Pi_{||}^\prime &=& \frac{3-\sqrt{3}}{6} \openone +
\frac{\sqrt{3}}{3}|\psi\rangle\langle \psi|,
\nonumber \\
\Pi_{\perp}^\prime &=& \frac{3-\sqrt{3}}{6} \openone +
\frac{\sqrt{3}}{3}|\psi_\perp\rangle\langle \psi_\perp|.
\label{PPRIME}
\end{eqnarray}
This POVM is symmetric (reflecting the symmetry of the program
$|\psi\rangle|\psi_\perp\rangle$). The fidelity
$F(\psi)$ is state independent and equal to the mean fidelity
\begin{equation}
F'=\frac{1}{2}\left(1+\frac{1}{\sqrt{3}}\right).
\label{FPERP}
\end{equation}
Notice that $F'={\cal{F}}_\perp$. This is not a mere coincidence,
the optimum strategy for program $|\psi\rangle|\psi_\perp\rangle$ is to
carry out an optimal estimation of $|\psi\rangle$ and then measure the
signal qubit in the basis formed by estimated state $|\psi_{\rm
est}\rangle$ and its orthogonal counterpart. The POVM (\ref{PPM}) is an
explicit implementation of this procedure. We emphasize here that
$F'$ is a maximum fidelity attainable with program $|\psi\rangle
|\psi_\perp\rangle$, because the corresponding CP map solves the
extremal equations (\ref{EXTREMALA}) and (\ref{EXTREMALB}). With the
program $|\psi\rangle|\psi\rangle$ we achieve the fidelity  $5/6\approx
0.8333$ which is higher than $F'$, hence the program
$|\psi\rangle|\psi\rangle$ exhibits better performance than
$|\psi\rangle|\psi_\perp\rangle$.

The aim of the measurement is to obtain some information on the system.
It can be illustrated on the following example: Alice encodes bits
of a message into quantum states and Bob tries to read the message making
measurements on these states. The average Bob's information (per bit) on
the message can be quantified as
\begin{equation}
    I = \sum_i r_i \left[ 1 + p_i \log_2 p_i
     + (1-p_i) \log_2 (1-p_i) \right],
 \label{AveInf}
\end{equation}
where $r_i$ is a fraction of bits that Bob knows with probability $p_i$
($\sum_i r_i = 1 $). It is reasonable to look for measurements that
maximize average information \cite{per}. However, as the average
information is a non-linear function of probabilities this is usually a
very difficult problem. We will not try to search for a CP map (or POVM)
that would maximize information instead of fidelity. Nevertheless, we
will calculate the average information obtainable by the measurements
described above that are optimal with respect to the fidelity. Let us
define fidelities $F_{||}=\langle \psi|\Pi_{||}|\psi\rangle$ and
$F_{\perp}=\langle\psi_\perp|\Pi_{\perp}|\psi_\perp\rangle$. If the
average rates of both the states are the same and equal to $1/2$
(generalization to an arbitrary rate is straightforward) then
\begin{eqnarray}
 & r_{||} = \left[ F_{||} + (1-F_{\perp}) \right]/2, \quad
 p_{||} = F_{||}/(2r_{||}), & \nonumber \\
 & r_{\perp} = \left[ F_{\perp} + (1-F_{||}) \right]/2, \quad
 p_{\perp} = F_{\perp}/(2r_{\perp}). &
\label{p+r}
\end{eqnarray}
For example, if $F_{||}=1$ and $F_{\perp}=1/2$ Bob gets result ``0''
always when Alice sends $|\psi\rangle$ and in half cases when she sends
$|\psi_\perp\rangle$. So, he knows the original bit with probability
$p_{||}=2/3$ and the average relative number of these results is
$r_{||}=3/4$. For result ``1'' the probability $p_{\perp}=1$ (this
result appears only if Alice sends $|\psi_\perp \rangle$) and
$r_{\perp}=1/4$. This leads to average information $I \doteq 0.311$. The
described situation corresponds to the above mentioned measurement with
the program state $| \psi \rangle$. Note that a symmetrized measurement
with the \emph{same} average fidelity ($F_{||}=F_{\perp}=3/4$) would
lead to \emph{lower} information ($I \doteq 0.189$).

For the case when the program state is $| \psi \rangle | \psi \rangle$
and $F_{||}=1$, $F_{\perp}=2/3$ the average information gets value $I
\doteq 0.459$. For the program state $| \psi \rangle | \psi_{\perp}
\rangle$, when $F_{||}=F_{\perp}=\frac{1}{2}(1 + \frac{1}{\sqrt{3}})$,
the information is $I \doteq 0.256$. Notice that this information is
\emph{lower} than the one obtained with the device programmed by the
single-qubit program state $| \psi \rangle$ ($0.256 < 0.311$).


\begin{figure}[t]
  \smallskip
  \centerline{\resizebox{0.85\hsize}{!}{\includegraphics*{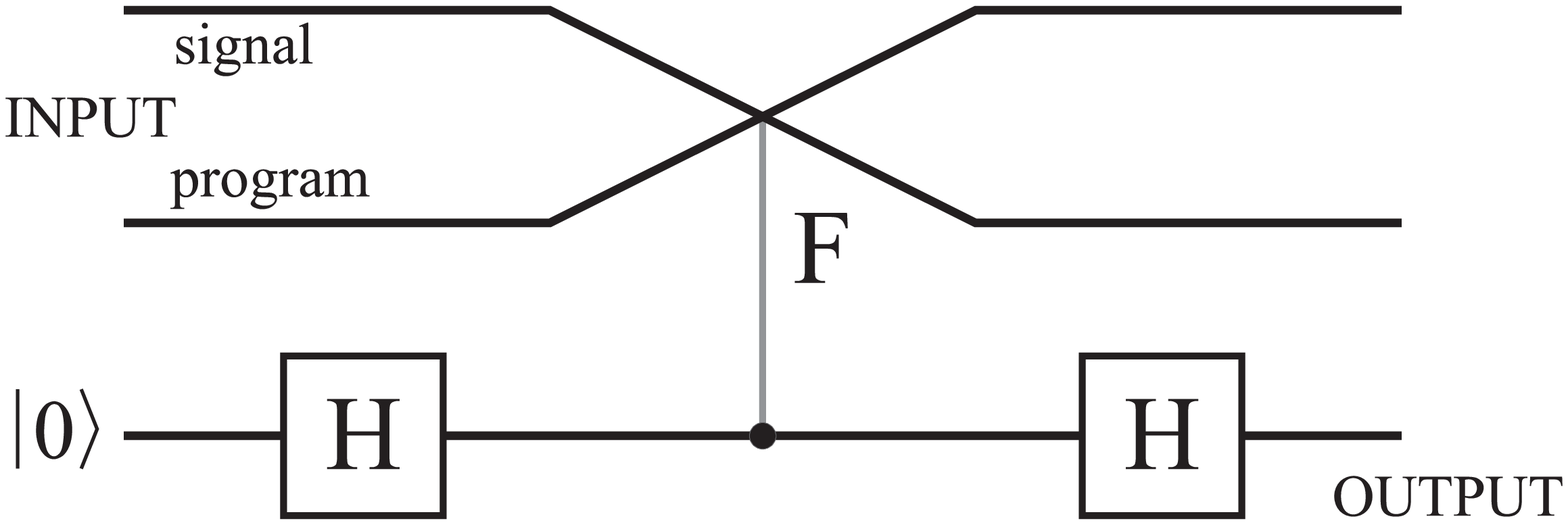}}}
  \smallskip
 \caption{Quantum circuit representing simple programmable
          multimeter: H -- Hadamard gate, F -- Fredkin gate.}
 \label{fig1}
\end{figure}

\begin{figure}[h]
  \smallskip
  \centerline{\resizebox{0.8\hsize}{!}{\includegraphics*{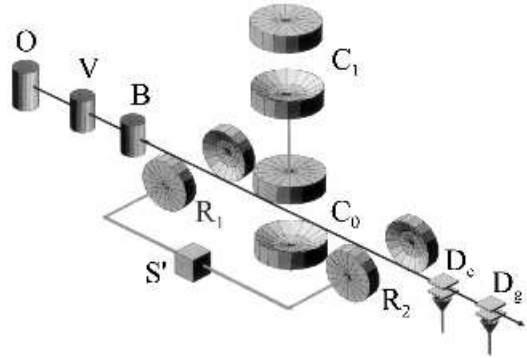}}}
  \smallskip
 \caption{Experimental realization of a programmable multimeter in cavity
          QED: O -- oven, V -- velocity selection, B -- excitation box,
          R$_{1}$, R$_{2}$ -- Ramsey zones, C$_{0}$, C$_{1}$ -- high-Q
          cavities, S$'$ -- microwave source, D$_{\rm e}$, D$_{\rm g}$
          -- ionization detectors.}
 \label{fig2}
\end{figure}

The proposed universal measurement devices can be, at least in
principle, realized experimentally. As an example, we describe now a
possible realization of the simplest one that uses a single-qubit
program register. Such a device can be built up from one Fredkin
(controlled swap) and two Hadamard gates as shown in Fig.~\ref{fig1} 
\cite{Ekert01}.
Signal and program enter the Fredkin gate as ``controlled'' qubits; the
result can be read out from the ancilla serving as a ``control'' qubit.
There are several ways how to implement the presented scheme on real
physical systems \cite{radim}. For example, we can employ an
experimental setup suggested previously in cavity-QED experiments
\cite{Raimond97} that is sketched in Fig.~\ref{fig2}. The Fredkin gate
is realized by two cavities C$_{0}$ and C$_{1}$ coupled by a waveguide
and it is controlled by an atomic ``probe'' affecting cavity field
through a nonlinear atom-field interaction in a large-detuning limit
\cite{Brune92}. This gate is inserted between two Hadamard
gates implemented as Ramsey atomic interferometers. The state of cavity
C$_{1}$ represents the signal, the state of cavity C$_{0}$ the program.
The program state can be set employing an atom-field interaction too. An
atom is prepared in the desired program state $a|{\rm g}\rangle +
b|{\rm e}\rangle$. This is done in zones B and R$_{1}$ (atoms come
from an oven O and only those with an appropriate velocity are selected
in zone V). An atom transition frequency must coincide with the cavity
frequency. Passing through the cavity C$_{0}$ the atom changes its
photonic state to $a|0\rangle+b|1\rangle$. Once the program and the
signal are encoded into the states of the cavities, 50:50 coupling
between C$_{0}$ and C$_{1}$ is established. Then the second atom,
initially prepared in the excited state, is sent through the Ramsey
interferometer. This atom is exposed to the nonlinear interaction with
the cavity field in a large-detuning limit (in  contrast to the first
one). Finally, the result of our programmed measurement process
encoded in the state of the atom can be revealed by the
field-ionization counters D$_{\rm e}$ end D$_{\rm g}$.


\begin{acknowledgments}

This research was supported under the project LN00A015 of the Ministry
of Education of the Czech Republic.

\end{acknowledgments}


\end{document}